\documentclass[]{eptcs}
\providecommand{\event}{TARK 2019} 
\usepackage{breakurl}             
\usepackage{underscore}           

\usepackage{amsthm, amsmath, amsfonts, amssymb}

\title{On the Consistency among Prior, Posteriors, and Information Sets (Extended Abstract)}
\author{Satoshi Fukuda
\institute{Department of Decision Sciences and IGIER\\
Bocconi University\thanks{I would like to thank Nobuo Koida, Massimo Marinacci, and Sujoy Mukerji for their insightful comments and discussions. I would also like to thank the TARK reviewers for their insightful comments.}\\
Milan, Italy}
\email{satoshi.fukuda@unibocconi.it}
}
\def\titlerunning{On the Consistency among Prior, Posteriors, and Information Sets}
\def\authorrunning{S. Fukuda}

\newtheorem{thm}{Theorem}
\newtheorem{prop}{Proposition}
\newtheorem{cor}{Corollary}

\theoremstyle{definition}

\newtheorem{exl}{Example}

\begin{document}
\maketitle

\begin{abstract}
This paper studies implications of the consistency conditions among prior, posteriors, and information sets on introspective properties of qualitative belief induced from information sets. The main result reformulates the consistency conditions as: (i) the information sets, without any assumption, almost surely form a partition; and (ii) the posterior at a state is equal to the Bayes conditional probability given the corresponding information set. Implications are as follows. First, each posterior is uniquely determined. Second, qualitative belief reduces to fully introspective knowledge in a \textquotedblleft standard'' environment. Thus, a care must be taken when one studies non-veridical belief or non-introspective knowledge. Third, an information partition compatible with the consistency conditions is uniquely determined by the posteriors. Fourth, qualitative and probability-one beliefs satisfy truth axiom almost surely. The paper also sheds light on how the additivity of the posteriors yields negative introspective properties of beliefs. 
\end{abstract}

\section{Introduction}

Agents in a strategic situation have two forms of beliefs. One is probabilistic beliefs, represented by a notion of types (\cite{Harsanyi}). The other is qualitative belief (or knowledge, if it is truthful), represented by information sets (\cite{Aumann76}). Consider, for instance, a dynamic game where each agent has knowledge about her past observations on the play, while she has probabilistic beliefs about her opponents' future plays. Qualitative belief plays a role when it comes to, say, studying consequences of common belief in rationality instead of common knowledge of rationality.\footnote{See, for instance, \cite{DekelGul} for the importance of capturing both knowledge and probabilistic beliefs. See, for example, \cite{Stalnaker94} for using qualitative and probabilistic beliefs for studying solution concepts of games.} These two kinds of beliefs are well studied in a rather separate manner, and somewhat surprisingly, little has been known about how reasoning based on one form of beliefs influences the other.

This paper examines introspective properties of qualitative belief induced by information sets from its relation with prior and posterior beliefs. First, I link prior and posteriors in a way that the prior probability coincides with the expectation of the posterior probabilities with respect to the prior. Second, I relate information sets and posteriors. An agent qualitatively believes (or knows) her own probabilistic beliefs. Also, if the agent qualitatively believes (or knows) something, then she believes it with probability one. I study how these linkages themselves yield introspective properties on qualitative beliefs.

Consider an agent, Alameda, who faces uncertainty about underlying states of the world. On the one hand, Alameda has a prior countably-additive probability measure. She also has a posterior probability measure at each realized state. These dictate her quantitative beliefs. While I analyze how the additivity of each posterior affects her reasoning, for now I assume each posterior to be countably additive. On the other hand, Alameda has a mapping, called a possibility correspondence. It associates, with each state, the set of states that she considers possible (the information set) at that state. I, as an analyst, derive properties of information sets, instead of directly assuming them. The framework is fairly parsimonious.

The main result (Theorem \ref{thm:main}) restates the consistency conditions as: (i) Alameda's information sets form a partition almost surely; and (ii) her posterior at each state coincides with the Bayes conditional probability given her corresponding information set. While information sets are usually exogenously assumed to form a partition, the main result demonstrates that the consistency conditions on the agent's qualitative and quantitative beliefs alone determine Bayes updating. 

While the main result has its own interest, I rather use it to derive the following four implications. The first implication (Corollary \ref{cor:uniquetype}) is that the consistency conditions uniquely determine the posterior at each state as the Bayes conditional probability given the associated information set. 

The second implication (Corollary \ref{cor:main}) is on the introspective properties of qualitative belief. Alameda's introspective abilities in her qualitative belief are reflected in properties of her information sets. When her information sets form a partition, her qualitative belief becomes knowledge (true belief) with full introspection. Truth axiom obtains: she can only know what is true. Her knowledge satisfies positive introspection: if she knows something then she knows that she knows it. Her knowledge also satisfies negative introspection: if she does not know something then she knows that she does not know it. 

Now, the second implication is that, in a standard countable model where the prior puts positive probability to every state, qualitative belief reduces to fully introspective knowledge. The consistency conditions alone determine this property, and thus qualitative belief reduces to fully introspective knowledge with no a priori assumption on qualitative belief. Also, the notions of qualitative and probability-one beliefs coincide, and each form of beliefs inherits the properties of the other form.  

Corollary \ref{cor:main} itself yields the following three additional implications. Its first implication  is on the evaluation of a solution concept of a game. If the analysts assume the consistency conditions among prior, posteriors, and information sets, qualitative belief reduces to knowledge even though the analysts would like to study, say, implications of common belief in rationality instead of common knowledge of rationality. That is, not only does qualitative belief reduce to knowledge for individual agents, but also common qualitative belief reduces to common knowledge (Corollary \ref{cor:ck}).

On a related point, second, if the analysts attempt to represent non-introspective knowledge violating negative introspection, then such non-introspective knowledge turns out to reduce to fully introspective knowledge. Such non-introspective knowledge is associated with unawareness: not only Alameda does not know an event $E$, but also she does not know that she does not know $E$ (\cite{MRTD, MRGEB}). Previous negative results on describing a non-trivial form of unawareness in a standard possibility correspondence model are based on some direct link between qualitative belief (knowledge) and unawareness (\cite{DLR, MRTD}). In contrast, Corollary \ref{cor:unawareness} demonstrates that a link between qualitative and probabilistic beliefs alone makes qualitative belief fully introspective knowledge and thus makes a standard state-space model with knowledge and probabilistic beliefs incapable of describing a non-trivial form of unawareness.

Third, still another implication of Corollary \ref{cor:main} is the violation of consistency between ex-ante and ex-post analyses in a non-partitional model. Suppose that the analysts aim to study non-introspective knowledge of a \textquotedblleft boundedly rational'' agent who is ignorant of her own ignorance. Assume that knowledge implies probability-one belief. Also, assume that the agent is positively introspective on her own probabilistic beliefs with respect to her knowledge (i.e., if she believes an event with probability at least $p$ (i.e., she $p$-believes the event), then she knows that she $p$-believes the event) in the same spirit as she has positive introspection on knowledge. Then, her prior and posteriors must violate the consistency condition, as the previous studies of non-partitional knowledge have demonstrated the inconsistency between ex-ante and ex-post evaluations of a decision problem.  

The third implication of the main result (Corollary \ref{cor:regular}) is on the uniqueness of an information partition (i.e., partitional information sets) compatible with the consistency conditions in a model featuring only probabilistic beliefs (i.e., prior and posteriors). This result justifies the use of the partition generated by the posteriors in the previous literature (e.g., \cite{BattigalliBonannoRIE, HalpernKBC, TanWerlangJET, VassilakisZamir}): if an agent is certain of her own posterior (\cite{Harsanyi, MertensZamir}), then, at each state, she has to be able to infer the set of states that generate the realized posterior. The resulting sets form the unique information partition. 

The fourth implication (Corollary \ref{cor:mainTA}) is that, while qualitative and probability-one beliefs may differ, both satisfy truth axiom almost surely.\footnote{For example, an agent believes with probability one (i.e., she is certain) that a random draw from $[0,1]$ is an irrational number but she does not know it (\cite{MondererSamet}). Recall, however, that Corollary \ref{cor:main} shows that, in a \textquotedblleft standard'' countable full-support environment, qualitative and probability-one beliefs coincide.} Common qualitative belief and common probability-one belief are also almost surely true. 

The paper also sheds light on the role of additivity in introspection (Propositions \ref{prop:pcertainty} and \ref{prop:operatorprop}). Now, suppose that the agent's posteriors are non-additive. First, if she does not $p$-believe an event $E$, then she may not be certain that she does not $p$-believe $E$. An agent with additive posteriors, however, would be certain of her own probabilistic ignorance. Second, additivity also implies negative introspection on probabilistic beliefs with respect to qualitative belief. If the agent does not $p$-believe an event then she may not qualitatively believe that she does not $p$-believe it. Again, an agent with additive posteriors would qualitatively believe her own probabilistic ignorance.

\subsection*{Related Literature}
This paper is related to the following four strands of literature: (i) derivation of Bayes updating from consistency between prior and posterior beliefs, (ii) interaction of knowledge and beliefs, (iii) non-partitional knowledge models, and (iv) the role of additivity in probabilistic reasoning. 

In the first strand of literature on Bayes updating, the main result (Theorem \ref{thm:main}) is closely related to \cite{Gaifman}, \cite{MertensZamir}, and \cite{SametBayesianism} in a purely probabilistic setting. In a single-agent perspective, these papers study how the consistency conditions between prior and posterior probabilities lead to Bayes updating.  Section \ref{sec:resultsmain} will discuss how these papers relate to the main result (Theorem \ref{thm:main}). As in \cite{MertensZamir}, the studies of existence of a common prior (e.g., \cite{BNCPA, Feinberg, GolubMorris, HeifetzCPA, HellmanGEB, MorrisECTA94, NehringET, SametCPAGEB}) also impose the consistency condition that the prior coincides with the expectation of the posteriors with respect to the prior.

The second is an extensive literature on the interaction between knowledge and beliefs in computer science, economics, game theory, logic, and philosophy. The consistency conditions between qualitative and probabilistic beliefs imposed in this paper are fairly common in economics and game theory in such contexts as epistemic characterizations of solution concepts for games, existence of a common prior, and canonical structures of agents' knowledge and beliefs (e.g., \cite{Aumann99, BattigalliBonannoRIE, DekelGul, MeierKB}). The validity of individual consistency conditions between knowledge and beliefs (i.e, \textquotedblleft knowledge entails beliefs'' and the knowledge of own beliefs) has been well studied since \cite{Hintikka} and \cite{Lenzen}. 

The third is on non-partitional knowledge that fails negative introspection. The studies of non-partitional structures include implications of common knowledge such as generalizations of Agreement theorem of \cite{Aumann76}, studies of solution concepts (\cite{BDGGEB, Geanakoplos, SametIgnorance, SametIJGT}),  foundations for information processing (errors) (\cite{Bacharach, MorrisJET, Shin}), and unawareness (\cite{DLR, MRTD, MRGEB}).\footnote{Propositions \ref{prop:pcertainty} and \ref{prop:operatorprop} suggest that non-additivity may yield the violation of probabilistic negative introspection: if the agent is not certain of an event, she may not be certain that she is not certain of it. The implication of such violation of probabilistic negative introspection would also be interesting.} See also \cite{DekelGul}.

Fourth, I turn to the role of additivity in probabilistic introspection. In the decision theory literature, such papers as \cite{GhirardatoET} and \cite{MukerjiET} study foundations for non-additive beliefs in terms of an agent's imperfect information processing. Although the framework of this paper is quite different from these decision-theoretic papers, these and this papers have the following similar intuition behind why the non-additivity is associated with the lack of introspection. An agent with non-additive beliefs  cannot imagine what the possible states are in her fullest extent. 


\section{An Epistemic Model}\label{sec:model}

\subsection{An Epistemic Model}\label{sec:epimodel}

This subsection formally defines a framework, which I call an epistemic model, for capturing quantitative and qualitative beliefs of an agent. For ease of exposition, unless otherwise stated, I restrict attention to a single agent model. 

An \textit{epistemic model} (a \textit{model}, for short) is a tuple $\overrightarrow{\Omega}:=\langle \Omega, \Sigma, \mu, P, t \rangle$. First, $\Omega$ is a non-empty set of \textit{states} of the world endowed with a $\sigma$-algebra $\Sigma$. Each element $E$ of $\Sigma$ is an \textit{event}. Denote the complement of an event $E$ by $E^{c}$ or $\neg E$. Second, $\mu: \Sigma \rightarrow [0,1]$ is a \textit{prior} countably-additive probability measure. Thus, $\langle \Omega, \Sigma, \mu \rangle$ forms a probability space. 

Third, $P: \Omega \rightarrow \Sigma$ is a \textit{possibility correspondence} with the measurability condition that $\{ \omega \in \Omega \mid P(\omega) \subseteq E \} \in \Sigma$ for each $E \in \Sigma$. It associates, with each state $\omega$, the set of states considered possible at that state. Note that $P(\omega)$ is assumed to be an event about which the agent herself reasons. The measurability condition will be used to introduce the qualitative belief operator. Thus, the possibility correspondence $P$ dictates the agent's qualitative belief on $\langle \Omega ,\Sigma \rangle$. Assume that $\mu$ and $P$ jointly satisfy $\mu(P(\cdot))>0$ as in standard partitional knowledge models (e.g., \cite{Aumann76}). 


Fourth, $t: \Omega \times \Sigma \rightarrow [0,1]$ is a \textit{type mapping} satisfying the following two measurability conditions. The type mapping $t$ dictates the agent's quantitative beliefs on $\langle \Omega, \Sigma \rangle$. The first assumption is: for each $E \in \Sigma$, the mapping $t (\cdot, E): \Omega \rightarrow [0,1]$ satisfies $(t(\cdot, E))^{-1}( [p,1]) = \{ \omega \in \Omega \mid t(\omega, E) \geq p \} \in \Sigma$ for all $p \in [0,1]$. That is, each $t(\cdot, E) : \langle \Omega, \Sigma \rangle \rightarrow \langle [0,1], \mathcal{B}_{[0,1]} \rangle$ is measurable with respect to the Borel $\sigma$-algebra $\mathcal{B}_{[0,1]}$ on $[0,1]$. This assumption allows the agent to reason about whether her degree of belief in an event $E$ is at least $p$. It will be used to define the agent's $p$-belief operators.  The second is: (i) $(\uparrow t(\omega)) :=\{ \omega' \in \Omega \mid t(\omega, \cdot) \leq t(\omega', \cdot) \} \in \Sigma$ and (ii) $(\downarrow t(\omega)) :=\{ \omega' \in \Omega \mid t(\omega', \cdot) \leq t(\omega, \cdot) \} \in \Sigma$ for all $\omega \in \Omega$. Note that $t(\omega, \cdot) \leq t(\omega', \cdot)$ means $t(\omega, E) \leq t(\omega', E)$ for all $E \in \Sigma$. I use similar abbreviations throughout the paper. If $\tilde{\omega} \in (\uparrow t(\omega))$, then $t(\tilde{\omega}, E)$ is always at least as high as $t(\omega, E)$ for any $E \in \Sigma$. 

For each $\omega \in \Omega$, call $t(\omega, \cdot)$ the \textit{type} at $\omega$. If a state $\omega$ realizes, the type $t(\omega, \cdot)$ at $\omega$ assigns, with each event $E$, the agent's quantitative (\textquotedblleft posterior'') belief in $E$. The idea behind the type mapping $t: \Omega \times \Sigma \rightarrow [0,1]$ is a Markov kernel  when each $t(\omega, \cdot)$ is a countably-additive probability measure (\cite{Gaifman, SametCPAGEB, SametJET00}). Here, each type $t(\omega, \cdot)$ is assumed to be a general set function. I do not assume any property of a set function on each type $t(\omega, \cdot)$ at this point, as I first study how each type inherits the properties of the prior $\mu$ by imposing the link between the prior $\mu$ and the type mapping $t$. 

Remarks on the second assumption are in order. Letting $[t(\cdot)]:= (\uparrow t(\cdot)) \cap (\downarrow t(\cdot))$, the set $[t(\omega)]$ consists of states $\tilde{\omega}$ indistinguishable from $\omega$ in that $t(\omega, \cdot) = t(\tilde{\omega}, \cdot)$. Intuitively, if the agent is perfectly certain of her quantitative beliefs, then, at each state $\omega$, she would be able to infer that the realization must be in $[t(\omega)]$. The second assumption ensures each $[t(\omega)]$ to be an object of the agent's beliefs.

While the measurability of $[t(\cdot)]$ is a standard assumption in a model in which an agent has countably-additive probabilistic beliefs, here instead I assume the measurability of $(\uparrow t(\cdot))$ and $(\downarrow t(\cdot))$ in order to see two different ways in which the agent can be certain of her quantitative beliefs later in Propositions \ref{prop:pcertainty} and \ref{prop:operatorprop}. Thus, before imposing consistency conditions on the agent's beliefs and thus studying a more-structured setting in which I derive main results, at this point I first delineate a general model to see how properties of beliefs (especially, additivity of quantitative beliefs) play a role in the agent's introspective reasoning. To see this point, by letting $[0,1]_{\mathbb{Q}} = [0,1] \cap \mathbb{Q}$, observe
\begin{align*}
(\uparrow t(\omega)) & = \bigcap_{(p,E) \in [0,1]_{\mathbb{Q}} \times \Sigma : t(\omega, E) \geq p} \{ \omega' \in \Omega \mid t(\omega', E) \geq p \} \text{ and}  \\
(\downarrow t(\omega)) & = \bigcap_{(p,E) \in [0,1]_{\mathbb{Q}} \times \Sigma : t(\omega, E) \leq p} \{ \omega' \in \Omega \mid t(\omega', E) \leq p \}.
\end{align*}
In the right-hand side of each expression, $p$ can also range over $[0,1]$ instead of $[0,1]_{\mathbb{Q}}$. Also, $\Sigma$ can be replaced with a countable algebra $\Sigma_{0}$ that generates $\Sigma$ if each $t(\omega, \cdot)$ is continuous with respect to both increasing and decreasing sequences of events and is monotone. 

When the agent reasons about $(\uparrow t(\cdot))$, she only uses her positive belief of the form, \textquotedblleft I believe an event with probability at least $p$.'' That is, when she does not believe an event $E$ with probability at least $p$, she does not take this information into account in inferring the true state. In contrast, when the agent reasons about $(\downarrow t(\cdot))$, she only uses her negative belief of the form, \textquotedblleft I do not believe an event with probability more than $p$.'' 

The distinction between $(\uparrow t(\cdot))$ and $[t(\cdot)]$ (and between $(\downarrow t(\cdot))$ and $[t(\cdot)]$) matters when the agent's belief is non-additive: if each $t(\omega, \cdot)$ is additive, then $(\uparrow t(\cdot)) = (\downarrow t(\cdot)) = [t(\cdot)]$. In this case, part (ii) of the second assumption is implied by part (i). In fact, the distinction between $(\uparrow t(\cdot))$ and $[t(\cdot)]$ is somewhat related to \cite{GhirardatoET} and \cite{MukerjiET} in the decision theory literature, which characterize non-additivity from the agent's \textquotedblleft perception'' (see also \cite{BonannoInfo} and \cite{LipmanSurvey}). To see this point, interpret $t$ as a mapping from $\Omega$ into the collection of set functions (with some given properties). On the one hand, $[t(\omega)]$ can be considered to be $t^{-1}(\{ t(\omega) \}):= \{ \tilde{\omega} \in \Omega \mid t(\omega)(\cdot) = t(\tilde{\omega})(\cdot) \}$. Thus, at state $\omega$, the agent is assumed to be able to observe a singleton $\{ t(\omega) \}$ so that she is able to infer that the true state is in $t^{-1}(\{ t(\omega) \})$. On the other hand, $(\uparrow t(\omega))$ can be regarded as $t^{-1}( \{ \mu \mid \mu(\cdot) \geq t(\omega)( \cdot) \} ):= \{ \tilde{\omega} \in \Omega \mid t(\tilde{\omega})(\cdot) \geq t(\omega)(\cdot) \}$. At state $\omega$, the agent is assumed to be able to observe a (generally) non-singleton set $\{ \mu \mid \mu(\cdot) \geq t(\omega)( \cdot) \}$. In \cite{GhirardatoET} and \cite{MukerjiET}, the agent has a limited observation on a non-empty set of consequences or signals (instead of own beliefs here) so that her beliefs may be non-additive. If, in contrast, she has a perfect observation on a singleton set of a consequence or a signal, then her beliefs are finitely additive. 

An epistemic model is a general framework for capturing the agent's quantitative and qualitative beliefs. In Section \ref{sec:boperators}, the agent's quantitative beliefs are represented as $p$-belief operators induced by the type mapping $t$, while her qualitative belief is represented by the qualitative belief operator induced by her possibility correspondence $P$. Section \ref{sec:relation} links (i) the prior $\mu$ and the type mapping $t$ and (ii) the possibility correspondence $P$ and the type mapping $t$.

\subsection{Quantitative and Qualittaive Belief Operators} \label{sec:boperators}

I introduce, in an epistemic model, quantitative and qualitative belief operators and their introspective properties. The agent's quantitative beliefs are captured by \textit{$p$-belief operators} (e.g., \cite{MondererSamet}). For each $(E,p) \in \Sigma \times [0,1]$, define $B^{p}(E):= \{ \omega \in \Omega \mid t(\omega, E) \geq p \} \in \Sigma$. The event $B^{p}(E)$ is the set of states at which the agent \textit{$p$-believes} $E$, i.e., she assigns \textquotedblleft probability'' at least $p$ to $E$. 

Two introspective properties of quantitative beliefs are in order. A model $\overrightarrow{\Omega}$ satisfies \textit{Certainty of ($p$-)Beliefs} if $t(\cdot, [t(\cdot)]) =1$ (\cite{Gaifman, MertensZamir, SametBayesianism, SametJET00}). Certainty of Beliefs states that the type at state $\omega$ puts probability one to the set of states indistinguishable from $\omega$ according to the type mapping $t$. To restate, if the agent has a perfect understanding of her own type mapping, she would be able to infer that, at each state $\omega$, the state is in $[t(\omega)]$ by unpacking the possible types.\footnote{\label{ft:certaintymu}Technically, \cite{Gaifman} and \cite{SametBayesianism} require $t(\omega, [t(\omega)]) =1$ $\mu$-almost surely.} If $\overrightarrow{\Omega}$ satisfies Certainty of Beliefs and if each $t(\omega, \cdot)$ is monotonic (i.e, $E \subseteq F$ implies $t(\omega,E) \leq t(\omega, F)$), then (i) $B^{p}(\cdot) \subseteq B^{1}B^{p}(\cdot)$ and (ii) $(\neg B^{p})(\cdot) \subseteq B^{1}(\neg B^{p})(\cdot)$. Part (i) states that if the agent $p$-believes an event $E$ then she $1$-believes that she $p$-believes $E$. Part (ii), on the other hand, states that if the agent does not $p$-believe an event $E$ then she $1$-believes that she does not $p$-believe $E$. Thus, Certainty of Beliefs implies full introspection in the above sense. Moreover, the idea of Certainty of Beliefs plays an important role in the construction of a universal Harsanyi type space (\cite{MertensZamir}).

If $\Sigma$ in a given model is generated by a countable algebra and if each $t(\omega, \cdot)$ is a countably-additive probability measure, then it follows from \cite[Theorem 3]{SametJET00} that the model  satisfies Certainty of Beliefs if and only if (hereafter, abbreviated as iff) $B^{p}(\cdot) \subseteq B^{1}B^{p}(\cdot)$.

Full introspection associated with Certainty of Beliefs comes from the fact that, at each state $\omega$, the agent always puts probability one to the set of states that are indistinguishable from $\omega$. I now disentangle the negative introspective property ($(\neg B^{p})(\cdot) \subseteq B^{1}(\neg B^{p})(\cdot)$) and Certainty of Beliefs. 

A model $\overrightarrow{\Omega}$ satisfies \textit{Positive Certainty of ($p$-)Beliefs} if $t(\cdot, (\uparrow t(\cdot))) =1$. At each state $\omega$, the agent puts probability one to the set of states $\tilde{\omega}$ with $t(\omega, \cdot) \leq t(\tilde{\omega}, \cdot)$. If the model satisfies Positive Certainty of Beliefs and if each $t(\omega, \cdot)$ is monotonic, then $B^{p}(\cdot) \subseteq B^{1}B^{p}(\cdot)$. Indeed, I show the sense in which Positive Certainty of Beliefs captures the positive introspective property $B^{p}(\cdot) \subseteq B^{1}B^{p}(\cdot)$.

\begin{prop}\label{prop:pcertainty}
Let a model $\overrightarrow{\Omega}$ satisfy the following: (i) $\Sigma$ is finite; (ii) each $t(\omega, \cdot)$ is monotonic; and (iii) if $t(\omega, E) = t(\omega, F) =1$ then $t(\omega, E \cap F)=1$. Then:
\begin{enumerate}
\item The model satisfies Positive Certainty of Beliefs iff $B^{p}(\cdot) \subseteq B^{1}B^{p}(\cdot)$.
\item Suppose further that $t(\cdot, \Omega) = 1$. Then, $t(\cdot, (\downarrow t(\cdot))) = 1$ iff $(\neg B^{p})(\cdot) \subseteq B^{1}(\neg B^{p})(\cdot)$.
\end{enumerate}
\end{prop}

Proposition \ref{prop:pcertainty} provides the sense in which Positive Certainty of Beliefs characterizes the positive introspective property $B^{p}(\cdot) \subseteq B^{1}B^{p}(\cdot)$, while (standard) Certainty of Beliefs implies both $B^{p}(\cdot) \subseteq B^{1}B^{p}(\cdot)$ and $(\neg B^{p})(\cdot) \subseteq B^{1}(\neg B^{p})(\cdot)$. 

Proposition \ref{prop:pcertainty} clarifies the role of additivity in introspection of quantitative beliefs. If each $t(\omega, \cdot)$ is additive then Certainty of Beliefs and Positive Certainty of Beliefs coincide as $(\uparrow t(\cdot)) = [t(\cdot)]$. Thus, under the setting of Proposition \ref{prop:pcertainty}, if each $t(\omega, \cdot)$ is additive, then the positive introspective property ($B^{p}(\cdot) \subseteq B^{1}B^{p}(\cdot)$) implies the negative introspective property ($(\neg B^{p})(\cdot) \subseteq B^{1}(\neg B^{p})(\cdot)$).\footnote{This remark applies to the aforementioned setting where $\Sigma$ is generated by a countable algebra and where each $t(\omega, \cdot)$ is a countably-additive probability measure.} 

I make two additional remarks on Proposition \ref{prop:pcertainty}. First, the condition (iii) is met, for instance, when a monotonic type $t(\omega, \cdot)$ is convex (i.e., $t(\omega, E) + t(\omega, F) \leq t(\omega, E \cap F) + t(\omega, E \cup F)$). That is, if $\Omega$ is finite and if each $t(\omega, \cdot)$ is a convex capacity then the assumptions in Proposition \ref{prop:pcertainty} are met. Second, I assume $\Sigma$ to be finite (and the condition (iii)) because an infinite $\sigma$-algebra $\Sigma$ is uncountable. Together with (i), the positive introspective property  $B^{p}(\cdot) \subseteq B^{1}B^{p}(\cdot)$ implies Positive Certainty of Beliefs. If $\Sigma$ is instead generated by a countable algebra and if each $t(\omega, \cdot)$ is continuous with respect to both increasing and decreasing sequences of events and is monotone, then the model satisfies Positive Certainty of Beliefs iff $B^{p}(\cdot) \subseteq B^{1}B^{p}(\cdot)$.

Next, I turn to the agent's qualitative belief. The possibility correspondence $P$ induces the \textit{qualitative belief operator} $K: \Sigma \rightarrow \Sigma$ defined by $K(E):= \{ \omega \in \Omega \mid P(\omega) \subseteq E \} \in \Sigma$ for each $E \in \Sigma$. The event $K(E)$ is the set of states at which the agent qualitatively believes $E$. Since $\omega \in K(P(\omega))$, it follows that $P (\omega) = \bigcap_{E \in \Sigma: \omega \in K(E)} E$. Hence, for a given qualitative belief operator $K$, the possibility correspondence $P$ that induces $K$ is unique (\cite{FukudaSetAlgebra}). Also, $K$ satisfies: (i) Monotonicity: $E \subseteq F$ implies $K(E) \subseteq K(F)$; (ii) (Countable) Conjunction: $\bigcap_{n \in \mathbb{N}} K (E_{n}) \subseteq K ( \bigcap_{n \in \mathbb{N}} E_{n} )$; and (iii) Necessitation: $K(\Omega) = \Omega$.

The following are well known (e.g., \cite{Aumann99, DekelGul, Geanakoplos, MorrisJET}). First, $K$ satisfies Truth Axiom ($K(E) \subseteq E$) iff $P$ is reflexive (i.e., $\omega \in P(\omega)$). Second, $K$ satisfies Positive Introspection ($K(\cdot) \subseteq K K (\cdot)$) iff $P$ is transitive (i.e., $\omega' \in P(\omega)$ implies $P(\omega') \subseteq P(\omega)$). Third, $K$ satisfies Negative Introspection ($(\neg K)(\cdot) \subseteq K (\neg K) (\cdot)$) iff $P$ is Euclidean (i.e., $\omega' \in P(\omega)$ implies $P(\omega) \subseteq P(\omega')$). Truth Axiom and Negative Introspection imply Positive Introspection. 

Note that no such introspective assumption on $P$ is imposed a priori. Thus, $K$ need not be the \textquotedblleft knowledge'' operator that satisfies Truth Axiom, although I denote the qualitative belief operator by $K$ to distinguish it from the $p$-belief operator $B^{p}$. Instead of assuming axioms on $P$, I derive properties of $P$ from how qualitative and quantitative beliefs interact with each other.

The domain of the qualitative belief operator $K$ is the $\sigma$-algebra $\Sigma$ because the collection of events $\Sigma$ describes entire objects of the agent's qualitative and quantitative beliefs. Technically, for any subset $A$ of $\Omega$, the set of states $\{ \omega \in \Omega \mid P(\omega) \subseteq A \}$ is a subset of $\Omega$. Thus, in principle, it is possible to define the qualitative belief operator from the entire power set of $\Omega$ into itself.\footnote{Objects of quantitative beliefs are also extended to all subsets of the states $\Omega$. Consider  the inner measure induced by a measure $\nu$: $\sup \{ \nu(E) \in [0,1] \mid E \in \Sigma \text{ and } E \subseteq A \}$ for any subset $A$ of $\Omega$.} However, specifying the domain $\Sigma$ determines depths of reasoning. Suppose that $\Sigma$ is not a $\sigma$-algebra but an algebra. For any finite length, the following form of event is well defined: the event that she qualitatively believes that she qualitatively believes that ... an event $E$ holds. Formally, $K^{n}(E) \in \Sigma$ for any $n \in \mathbb{N}$. The intersection of all these events $\bigcap_{n \in \mathbb{N}} K^{n}(E)$, however, may not be well defined. In this sense, assigning the domain $\Sigma$ as an algebra provides the agent with finite depths of reasoning. Likewise, if $\Sigma$ is a $\sigma$-algebra, the agent can reason about her countable depths of reasoning. But the conjunction of an uncountable number of events may not be an event. Indeed, the existence and non-existence of a canonical (\textquotedblleft universal'' or \textquotedblleft terminal'') interactive type/belief/knowledge space hinges on such specification of domain (e.g., \cite{FukudaUQB, HeifetzSametKnowledge, MeierNonExistence, MeierKB}). Thus, I take the domain of the qualitative belief operator $K$ to be the $\sigma$-algebra $\Sigma$.

\subsection{Relations among Prior, Posteriors, and Information Sets}\label{sec:relation}

The previous section introduced the agent's quantitative and qualitative beliefs from her type mapping and possibility correspondence, respectively. Now, I relate (i) the prior and the type mapping and (ii) the type mapping and the possibility correspondence (i.e., quantitative and qualitative beliefs). Throughout this subsection, fix a model $\overrightarrow{\Omega}$.

First, assume that the prior and the type mapping jointly satisfy the invariance condition that the prior probability of an event $E$ coincides with the expectation of the posteriors of $E$ with respect to the prior. Formally, the model satisfies \textit{Invariance} if 
\begin{equation*}
\mu (\cdot) = \int_{\Omega} t(\omega, \cdot) \mu (d\omega).
\end{equation*}

This consistency condition is especially used to characterize (existence of) a common prior as discussed in the introduction. Note also that, in accordance with this literature, one can define the prior $\mu$ as a (countably-additive) probability measure that satisfies the Invariance condition. \cite{HMSbelief} also impose this condition in their probabilistic model of unawareness.

Second, I introduce two introspective properties that relate qualitative and quantitative beliefs.  The model $\overrightarrow{\Omega}$ satisfies \textit{Entailment} if $t(\cdot, P(\cdot)) = 1$. Entailment states that, at each state, the agent assigns probability one to the set of states that she considers possible at that state. Since $\omega \in K(P(\omega))$, if each type $t(\omega, \cdot)$ is monotonic then Entailment is expressed, in terms of operators, as the stronger condition $K(\cdot) \subseteq B^{1}(\cdot)$. If qualitative belief reduces to knowledge, then Entailment states that knowledge entails probability-one belief (see, e.g., \cite{BattigalliBonannoRIE, DekelGul, Hintikka}).

Next, $\overrightarrow{\Omega}$ satisfies \textit{Self-Evidence of ($p$-)Beliefs} if $\omega' \in P(\omega)$ implies $t(\omega, \cdot) \leq t(\omega', \cdot)$, i.e., $P(\omega) \subseteq (\uparrow t(\omega))$. It provides a consistency requirement between the possibility correspondence $P$ and the type mapping $t$: if the agent considers $\omega'$ possible at $\omega$, then, as long as she assigns probability at least $p$ to an event $E$ at $\omega$, she also assigns probability at least $p$ to $E$ at $\omega'$. If each $t(\omega, \cdot)$ is additive, then Self-Evidence of Beliefs is re-written as $P(\omega) \subseteq [t(\omega)]$ (i.e., if $\omega' \in P(\omega)$ then $t(\omega, \cdot) = t(\omega', \cdot)$). If qualitative belief reduces to knowledge, then the latter condition ($P(\omega) \subseteq [t(\omega)]$) is commonly imposed in economics and game theory.\footnote{In philosophy, \cite[Section 3.7]{Hintikka} rejects Self-Evidence of Beliefs. See also \cite[Chapter 4]{Lenzen} for critical assessments of the rejection of this axiom.} I, however, formulate Self-Evidence of beliefs so as to examine the effect of the additivity of types. The following proposition shows that Self-Evidence of Beliefs captures positive introspection. 

\begin{prop}\label{prop:operatorprop}
\begin{enumerate}
\item The model satisfies Self-Evidence of Beliefs iff $B^{p}(\cdot) \subseteq K (B^{p})(\cdot)$. 
\item $P(\cdot) \subseteq (\downarrow t(\cdot))$ iff $(\neg B^{p})(\cdot) \subseteq K ( \neg B^{p})(\cdot)$. 
\end{enumerate}
\end{prop}

Proposition \ref{prop:operatorprop} states that Self-Evidence of Beliefs means: whenever the agent $p$-believes an event $E$, she qualitatively believes (knows) that she $p$-believes $E$.\footnote{Note that I do not impose the following form of introspection: $B^{p}(E) \subseteq B^{p}K(E)$ (i.e., whenever the agent $p$-believes an event $E$, she $p$-believes that she qualitatively believes $E$). This form of interaction between \textquotedblleft knowledge and belief (certainty)'' is considered in computer science, logic, and philosophy (see, \cite{HalpernJPL, Lenzen} and the references therein).} Proposition \ref{prop:operatorprop} implies that $P(\cdot) \subseteq [t(\cdot)]$ iff $B^{p}(\cdot) \subseteq K (B^{p})(\cdot)$ and $(\neg B^{p})(\cdot) \subseteq K ( \neg B^{p})(\cdot)$. Proposition \ref{prop:operatorprop} again disentangles the role of the additivity of types by considering $(\uparrow t(\cdot))$ and $(\downarrow t(\cdot))$.

Two remarks are in order. First, Entailment and Self-Evidence of Beliefs imply Positive Certainty of Beliefs, provided that each $t(\omega, \cdot)$ is monotonic. Second, suppose Invariance and Self-Evidence of Beliefs. Using the assumption made in Section \ref{sec:epimodel} that $\mu(P(\cdot))>0$ as in a standard possibility correspondence model, $\mu(E)=0$ iff $t(\omega, E)=0$ for all $\omega \in \Omega$. I say that $E \subseteq F$ $\mu$-almost surely ($\mu$-a.s.) if $\mu(E \setminus F) = 0$. The above argument implies the consistency condition that $E \subseteq F$ $\mu$-a.s. iff $E \subseteq F$ $t(\omega, \cdot)$-a.s. for all $\omega \in \Omega$, under Invariance and Self-Evidence of Beliefs. Likewise, I say that $E = F$ $\mu$-almost surely if $\mu(E \triangle F) = 0$, where $\triangle$ stands for the symmetric difference.

With these definitions in mind, the main result (Theorem \ref{thm:main}) characterizes conditions on a given model under which the agent's type at each state coincides with the Bayes conditional probability given her information set at that state. To that end, call the model $\overrightarrow{\Omega}$ \textit{regular} if each $t(\omega, \cdot)$ is a countably-additive probability measure and if the model satisfies Invariance (the consistency condition between the prior and the type mapping), Entailment and Self-Evidence of Beliefs ($t(\cdot, P(\cdot))=1$ and $P(\cdot) \subseteq (\uparrow t(\cdot))$, which are the consistency conditions between qualitative and quantitative beliefs). Note that, since the agent's types are additive in a regular model, Positive Certainty of Beliefs reduces to Certainty of Beliefs.   

\section{Results}\label{sec:results}

\subsection{Main Result and Uniqueness of the Type Mapping}\label{sec:resultsmain}

I present the main result that fully characterizes a regular model. A model is regular iff (i) each type $t(\omega, \cdot)$ is derived, through Bayes updating, from the prior $\mu$ conditional on the information set $P(\omega)$; and (ii) the information sets form a partition almost surely.  

\begin{thm}\label{thm:main}
A model $\overrightarrow{\Omega}$ is regular iff (i) $t(\cdot, \cdot) = \mu (\cdot \mid P(\cdot))$, (ii) $P(\cdot) \subseteq [t(\cdot)]$, and (iii) $P(\cdot) \supseteq [t(\cdot)]$ $\mu$-almost surely. 
\end{thm}

Five remarks on Theorem \ref{thm:main} are in order. First, one can require each $t(\omega, \cdot)$ only to be finitely additive in a regular model. The theorem does not hinge on the continuity of each $t_{i}(\omega, \cdot)$. Indeed, each finitely-additive type $t(\omega, \cdot)$ becomes countably additive as it inherits countable additivity from the prior $\mu$ by part (i). Second, suppose that $\mu$ and every $t(\omega, \cdot)$ are only finitely additive while keeping all the other assumptions (note that the integral of a bounded measurable mapping with respect to $\mu$ is well defined). The \textquotedblleft only if'' part still holds. That is, the consistency conditions still imply that each type $t(\omega, \cdot)$ is the Bayes conditional probability $\mu (\cdot \mid P(\cdot))$ and that the collection of information sets $\{ P(\omega) \}_{\omega \in \Omega}$ almost surely forms a partition. The \textquotedblleft if'' part holds when $\{ [t(\cdot)] \}$ forms a finite partition.

Third, relax the assumption $\mu(P(\cdot))>0$ for a regular model $\overrightarrow{\Omega}$. Then, any regular model satisfies $\mu(E \cap P(\omega)) =  \mu(P(\omega)) t(\omega,E)$ for all $(\omega,E) \in \Omega \times \Sigma$ as well as (ii) and (iii).  

Fourth, it is not trivial at all that $P(\cdot) = [t(\cdot)]$ ($\mu$-almost surely). While both objects represent the agent's information at each state, $P(\omega) = \bigcap_{E \in \Sigma: \omega \in K(E)} E$ and $[t(\omega)] = \bigcap_{(E,p) \in \Sigma \times [0,1]: \omega \in B^{p}(E)} B^{p}(E)$ in a regular model.  

Fifth, since part (iii) only requires $P(\cdot) \supseteq [t(\cdot)]$ $\mu$-almost surely, $\{ P(\omega) \}_{\omega \in \Omega}$ may fail to be a partition. Thus, the qualitative belief operator $K$ may violate Truth Axiom ($K(E) \subseteq E$), Positive Introspection ($K(\cdot) \subseteq KK(\cdot)$), and Negative Introspection ($(\neg K)(\cdot) \subseteq K(\neg K)(\cdot)$). The model, however, satisfies introspection of the form $B^{p}(\cdot) \subseteq KB^{p}(\cdot)$ and $(\neg B^{p})(\cdot) \subseteq K(\neg B^{p})(\cdot)$. Section \ref{sec:partition} examines a special case where $\Omega$ is countable, $\Sigma=2^{\Omega}$, and where $\mu(\{ \cdot \})>0$. There, $P(\cdot) = [t(\cdot)]$ forms a partition, that is, the agent's qualitative belief satisfies Truth Axiom, Positive Introspection, and Negative Introspection. Moreover, qualitative belief and probability-one belief coincide as $K=B^{1}$. Section \ref{sec:TA} compares the qualitative belief and probability-one belief operators in detail.

Sixth, in part (iii), if one considers a different notion of $\mu$-almost-sureness measuring the set of $\omega$ such that $P(\omega) \supseteq [t(\omega)]$, it may be the case that the $\mu$-measure of such a set may not be equal to $1$, provided that such a set is measurable (Example \ref{ex:introspection} in Appendix \ref{sec:appendix} is such an example). Since $P(\omega)$ dictates the agent's qualitative belief at state $\omega$, I consider, at each state $\omega$, whether $P(\cdot) \supseteq [t(\cdot)]$ holds in a certain sense. Part (iii) says that $P(\omega) \supseteq [t(\omega)]$ holds $\mu$-almost surely at each given state $\omega \in \Omega$.

Theorem \ref{thm:main} relates to the following previous results on the consistency conditions between prior and posteriors where quantitative beliefs are sole primitives. First, \cite{MertensZamir} ask when an agent's posterior beliefs are derived from her (or common) prior conditional on her information. \cite[Proposition 4.2]{MertensZamir} show that if a given quantitative belief model $\langle \Omega, \Sigma, \mu, t \rangle$ satisfies Invariance and Certainty of Beliefs then the agent's type $t(\omega, E)$ turns out to be the Bayes conditional probability $\mu(E \mid [t(\omega)])$ whenever it is well defined.\footnote{In the similar way to the proof of Theorem \ref{thm:main}, the following can be established. For a given model $\overrightarrow{\Omega}$, let each $t(\omega, \cdot)$ be countably additive. Then, the model satisfies Invariance and Certainty of Beliefs iff $t(\cdot, \cdot) = \mu(\cdot \mid [t(\cdot)])$ whenever the right-hand side is well defined.} 

Second, \cite{SametBayesianism} calls a quantitative belief model $\langle \Omega, \Sigma, \mu, t \rangle$ to be Bayesian if it satisfies Certainty of Beliefs ($\mu$-almost surely) and Invariance. \cite{Gaifman} and \cite{SametBayesianism} characterize a Bayesian model by the consistency requirement that the prior conditioned on some specification of the posterior beliefs must agree with the specification.

These results and Theorem \ref{thm:main} of this paper \textit{derive} Bayes updating from epistemic properties within a model. Theorem \ref{thm:main} states that, in an environment in which an agent's quantitative and qualitative beliefs are both present, the interaction between prior and posteriors (i.e., Invariance) and the ones between posteriors and possibility correspondence (i.e., Entailment and Self-Evidence of Beliefs) give rise to Bayes updating within the model, provided each type is countably (indeed, finitely) additive. 

From now on, I examine the implications of Theorem \ref{thm:main}. The first immediate corollary is the uniqueness of a type mapping. Not only does Theorem \ref{thm:main} justify Bayes updating conditional on information sets, but also it implies that if a model is regular then the type mapping is uniquely determined by the other two ingredients of the model (namely, the prior $\mu$ and the possibility correspondence $P$) through Bayes conditional probabilities. 

\begin{cor}\label{cor:uniquetype}
Let $\langle \Omega, \Sigma, \mu, P, t \rangle$ and $\langle \Omega, \Sigma, \mu, P,  t' \rangle$ be regular models (where $\Omega, \Sigma, \mu$, and $P$ are common). Then, $t=t'$.  
\end{cor}

Similarly, a possibility correspondence is almost surely unique: if $\langle \Omega, \Sigma, \mu, P, t \rangle$ and $\langle \Omega, \Sigma, \mu, P', t \rangle$ are regular models (where $\Omega, \Sigma, \mu$, and $t$ are common), then, for each $\omega \in \Omega$, $P(\omega)=P'(\omega)$ $\mu$-almost surely. While possibility correspondences are unique $\mu$-almost surely, the resulting qualitative belief operators may not necessarily satisfy $K=K'$ $\mu$-almost surely (Examples \ref{ex:boperators} and \ref{ex:introspection} in Appendix \ref{sec:appendix} can be seen as such examples). Section \ref{sec:partition} shows that if $P$ in a regular model forms a partition then $P$ is unique.

\subsection{Partitional Properties of Qualitative Belief}\label{sec:partition}

Call a model $\overrightarrow{\Omega}$ \textit{discrete} if $\Omega$ is countable, $\Sigma = 2^{\Omega}$, and if $\mu(\{ \cdot \}) >0$.\footnote{In a discrete model, $\Sigma = 2^{\Omega}$ is generated by a countable algebra, e.g., $\Sigma_{0} = \{ E \in 2^{\Omega} \mid E \text{ is finite or } E^{c} \text{ is finite} \}$.} Under this \textquotedblleft standard'' setting, the following corollary demonstrates that qualitative belief necessarily becomes knowledge and that probability-one belief and knowledge coincide.

\begin{cor}\label{cor:main}
\begin{enumerate}
\item A discrete model $\overrightarrow{\Omega}$ is regular iff (i) $P(\cdot) = [t(\cdot)]$ and (ii) $t(\cdot, \cdot) = \mu (\cdot \mid P(\cdot))$. In this case, $K=B^{1}$ (satisfies Truth Axiom, Positive Introspection, and Negative Introspection). 

\item Let $\overrightarrow{\Omega}$ be a regular model with $\{ \omega \} \in \Sigma$ for all $\omega \in \Omega$. The model is discrete iff $B^{1}$ satisfies Truth Axiom. In this case, $K=B^{1}$ (satisfies Truth Axiom, Positive Introspection, and Negative Introspection). 
\end{enumerate}
\end{cor}

In a regular discrete model, the possibility correspondence $P(\cdot)$ exactly coincides with $[t(\cdot)]$, and knowledge and certainty (probability-one belief) coincide with each other, irrespective of assumptions on information sets. Since $P(\cdot) = [t(\cdot)]$ forms a partition on the state space, $K=B^{1}$ satisfies Truth Axiom, Positive Introspection, Negative Introspection, (Countable) Conjunction, Monotonicity, and Necessitation. Moreover, $B^{1} = K$ satisfies the following form of strong conjunction property. For any collection of events $\mathcal{E} \in \mathcal{P}(\Sigma)$ with $\bigcap \mathcal{E} \in \Sigma$, $\bigcap_{E \in \mathcal{E}} B^{1}(E) \subseteq B^{1}(\bigcap \mathcal{E})$.

Moreover, Corollary \ref{cor:main} implies that possibility coincides with assigning positive probability in the sense that $P(\omega) = \{ \omega' \in \Omega \mid t(\omega, \{ \omega' \}) >0 \}$ for all $\omega \in \Omega$.\footnote{\cite{HalpernKBC} studies certainty (probability-one belief) by defining the \textquotedblleft support relation'' between two states by $t(\omega, \{ \omega' \}) >0$. \cite{SametCPAGEB} studies the (Markov transition) matrix generated by $(t(\omega, \{ \omega' \}))_{\omega, \omega' \in \Omega}$. \cite{MorrisJET} derives qualitative belief from preferences, and under certain condition, the notion of possibility reduces to assigning positive probability.} In a quantitative belief model, an information set at a state is often defined as the support of the type at that state. Section \ref{sec:uniquepartition} demonstrates that if $\{ P(\omega)\}_{\omega \in \Omega}$ forms a partition then it is uniquely determined by $P(\cdot) = [t(\cdot)]$.  

Corollary \ref{cor:main} suggests that a care must be taken of the agent's qualitative belief if the analysts study an epistemic characterization of a solution concept for a game. One of the most standard ways to study an epistemic characterization of a solution concept for a game is to characterize an agent's beliefs by a prior, posteriors, and a possibility correspondence on a state space. If the possibility correspondence is partitional, then the model captures fully introspective knowledge and probabilistic beliefs. As discussed in the introduction, for example, such a model can capture an agent's fully introspective knowledge about her past observations and her beliefs about her opponents' future plays in a dynamic game.\footnote{Another approach is a (product) type space approach. See, for example, \cite{BattigalliBonannoRIE, DekelGul, DekelSiniscalchiHandbook, Stalnaker94}, and the references therein for epistemic characterizations of solution concepts.}

I examine the implications of Corollary \ref{cor:main} when the information sets do not form a partition. Suppose that the analysts introduce qualitative belief instead of knowledge when they study, for example, implications of common belief in rationality instead of common knowledge of rationality (see, e.g., \cite{BonannoHandbook15, BonannoTsakas, Stalnaker94}). Qualitative belief violates Truth Axiom if the information sets fail reflexivity. Corollary \ref{cor:main} points to the importance of figuring out the relations between prior, posteriors, and information sets in a discrete model because qualitative belief reduces to (fully introspective) knowledge despite the analysts' purpose. Section \ref{sec:CB} formally introduces notions of common beliefs. 

Next, a violation of fully introspective knowledge may also come from lack of introspection. Thus, suppose that the analysts would like to study knowledge of a (single) agent who violates Negative Introspection: the agent does not know an event $E$ at a state, and she does not know that she does not know it at that state. In other words, consider unawareness of an agent in a state space model with knowledge and probabilistic beliefs. In a discrete regular model, since $K$ has to satisfy Negative Introspection, there is no state at which the agent is unaware of any event $E$ (i.e., she does not know $E$ and she does not know that she does not know $E$). I formulate it as an immediate corollary.

\begin{cor}\label{cor:unawareness}
In a discrete regular model $\overrightarrow{\Omega}$,  the agent is unaware of nothing: $(\neg K)(\cdot) \cap (\neg K)^{2}(\cdot) = \emptyset$.
\end{cor} 

I discuss two points that the corollary makes on unawareness in standard state space models. First, the previous negative results on the possibility of describing a richer form of unawareness (e.g., \cite{DLR, MRTD}) impose some direct links between knowledge and unawareness. Corollary \ref{cor:unawareness} suggests that, once an agent possesses her quantitative belief, there is a new channel through which the agent can have fully introspective knowledge and thus she is fully aware of everything. 

Second, when it comes to representing non-introspective knowledge in a probabilistic environment, additivity of posteriors and the invariance condition would be strong so that knowledge eventually becomes fully introspective. To see this point, suppose $P(\cdot) \neq [t(\cdot)]$ due to the failure of Negative Introspection (of $K$) in a discrete model as above. I consider non-partitional models in broader contexts, as the same conclusion as above can already be drawn with respect to solution concepts of games and the implications of common knowledge such as the Agreement theorem (\cite{Aumann76}) in non-partitional models. Now, the model has to violate either Invariance, Entailment, Self-Evidence of Beliefs, or the assumption that each type $t(\omega, \cdot)$ is additive. 

Since the information sets represent the agent's knowledge, assume Entailment: if she knows an event then she $1$-believes the event. Also, assume Self-Evidence of Beliefs. Just as the agent's knowledge is positively introspective, if she $p$-believes an event $E$ then she knows that she $p$-believes $E$. Thus, if each type $t(\omega, \cdot)$ is additive, then Corollary \ref{cor:main} implies that the model has to violate Invariance. Likewise, under Invariance, some type $t(\omega, \cdot)$ may be non-additive. 

Consider a discrete model $\overrightarrow{\Omega} = \langle \Omega, \Sigma, \mu, P, t \rangle$ with $P(\cdot) \neq [t(\cdot)]$ and $t(\cdot, P(\cdot)) = 1$ (i.e., Entailment). The type at each state $\omega$ coincides with the Bayes conditional probability $t(\omega,E) = \frac{\mu(E \cap P(\omega))}{\mu(P(\omega))}$ for all $(\omega, E)$ iff the likelihood ratio between types coincides with the likelihood ratio between prior probabilities given an information set $\frac{t(\omega, E)}{t(\omega,F)} = \frac{\mu(E \cap P(\omega))}{\mu(F \cap P(\omega))}$ for all $(\omega, E, F)$ with $\mu(F \cap P(\omega))>0$ and $t(\omega,F)>0$. If this is the case, then each type $t(\omega, \cdot)$ is  additive. If the model satisfies Self-Evidence of Beliefs, then it has to violate Invariance. 

This observation sheds light on the comparison between ex-ante and ex-post analyses or the value of information in non-partitional knowledge models as discussed by, for example, \cite{DekelGul} and \cite{Geanakoplos}. This observation provides an intuition behind why \textquotedblleft dynamic inconsistency'' occurs in a non-partitional (i.e., reflexive and transitive) environment.\footnote{\cite{Gaifman} also discusses the violation of a Bayesian belief model $\langle \Omega, \Sigma, \mu, t \rangle$ in terms of \textquotedblleft dynamic inconsistency'' between ex-ante and ex-post analyses.} This observation also sheds light on quantitative belief updating in a non-partitional environment. In the literature studying non-partitional information sets, it has been assumed that an agent updates her probabilistic assessment according to the Bayes rule given a non-partitional information set as in the previous argument. As \cite[p.147]{DekelGul} put it, however, \textquotedblleft there are essentially no results that justify stapling traditional frameworks together with non-partitions.''\footnote{The probabilistic approach to unawareness by \cite{HMSbelief} considers an extended structure consisting of multiple sub-state-spaces. While an agent's beliefs satisfy Invariance within each sub-space, she exhibits unawareness in the entire enriched model.}

\subsection{Common Qualitative and Quantitative Beliefs} \label{sec:CB}

As discussed in the previous subsection, I define an interactive epistemic model to introduce notions of common beliefs. Let $I$ be a non-empty at-most-countable set of agents. An \textit{interactive epistemic model} is $\langle (\Omega, \Sigma, \mu), (P_{i}, t_{i})_{i \in I} \rangle$ such that $\langle \Omega, \Sigma, \mu, P_{i}, t_{i} \rangle$ is a model for each agent $i \in I$. Denote by $B^{p}_{i}$ and $K_{i}$ agent $i$'s $p$-belief operator and qualitative belief operator, respectively. Call the interactive epistemic model $\langle (\Omega, \Sigma, \mu), (P_{i}, t_{i})_{i \in I} \rangle$ \textit{regular} if $\langle \Omega, \Sigma, \mu, P_{i}, t_{i} \rangle$ is regular for each $i \in I$. Likewise, call the interactive epistemic model \textit{discrete} if $\Omega$ is countable, $\Sigma = 2^{\Omega}$, and if $\mu(\{ \cdot \}) >0$. For a regular model, define the (iterative) common $p$-belief operator $C^{p}: \Sigma \rightarrow \Sigma$ as $C^{p}(\cdot):=\bigcap_{n \in \mathbb{N}} (\bigcap_{i \in I} B^{p}_{i})^{n}(\cdot)$. Likewise, define the (iterative) common qualitative belief operator $C: \Sigma \rightarrow \Sigma$ by $C(\cdot):=\bigcap_{n \in \mathbb{N}} (\bigcap_{i \in I} K_{i})^{n}(\cdot)$.\footnote{Call an event $E$ a common $p$-basis if $E \subseteq \bigcap_{i \in I} B^{p}_{i}(F)$ for any $F \in \Sigma$ with $E \subseteq F$ (\cite{FukudaCB}). In an interactive epistemic model, an event $E$ is \textit{common $p$-belief} at $\omega$ if there is a common $p$-basis $F$ with $\omega \in F \subseteq \bigcap_{i \in I} B^{p}_{i}(E)$. It can be seen that, in any regular interactive epistemic model, $E$ is common $p$-belief at $\omega$ iff $\omega \in C^{p}(E)$. Likewise, an event $E$ is a common basis if $E \subseteq \bigcap_{i \in I} K_{i}(F)$ for any $F \in \Sigma$ with $E \subseteq F$. In an interactive epistemic model, an event $E$ is \textit{common qualitative belief} at $\omega$ if there is a common basis $F$ with $\omega \in F \subseteq \bigcap_{i \in I} K_{i}(E)$. It can be seen that, in any interactive epistemic model, $E$ is common qualitative belief at $\omega$ iff $\omega \in C(E)$ (\cite{FukudaCB}). If $E$ is common qualitative belief at $\omega$, then $E$ is common $p$-belief at $\omega$.} The common qualitative belief operator is induced by the transitive closure of $(P_{i})_{i \in I}$.

\begin{cor}\label{cor:ck}
In a discrete regular interactive epistemic model $\langle (\Omega, \Sigma, \mu), (P_{i}, t_{i})_{i \in I} \rangle$, common qualitative belief and common $1$-belief coincide: $C=C^{1}$. Moreover, $C=C^{1}$ satisfies Truth Axiom, Positive Introspection, and Negative Introspection. That is, common qualitative belief reduces to common knowledge. 
\end{cor} 

In a discrete regular interactive epistemic model, each agent's possibility correspondence is  partitional so that the Agreement theorem (\cite{Aumann76}) holds. More generally, I establish the following Agreement theorem(s) (\cite{Aumann76, MondererSamet, NeemanAgree}) for a regular interactive epistemic model. The result holds as long as there exists some prior $\mu$ such that $\langle (\Omega, \Sigma, \mu), (P_{i}, t_{i})_{i \in I} \rangle$ is regular. 

\begin{prop}\label{prop:agreement}
Let $\langle (\Omega, \Sigma, \mu), (P_{i}, t_{i})_{i \in I} \rangle$ be regular. If $C^{p}( \bigcap_{i \in I}\{ \tilde{\omega} \in \Omega \mid t_{i}(\tilde{\omega}, E) = r_{i} \}) \neq \emptyset$, then $|r_{i} - r_{j}| \leq 1-p$ for all $i,j \in I$. Especially, if $C( \bigcap_{i \in I} \{ \tilde{\omega} \in \Omega \mid t_{i}(\tilde{\omega}, E) = r_{i} \}) \neq \emptyset$, then $r_{i} = r_{j}$ for all $i,j \in I$.
\end{prop}

Invariance plays an important role in the proposition, especially when the assumption $\mu(P_{i}(\cdot))>0$ is dropped. In this general case, If $\mu( C^{p}( \bigcap_{i \in I}\{ \tilde{\omega} \in \Omega \mid t_{i}(\tilde{\omega}, E) = r_{i} \}) ) >0$, then $|r_{i} - r_{j}| \leq 1-p$ for all $i,j \in I$. On the one hand, by invariance, the prior probability in $E$ conditional on $i$'s $p$-belief in the common $p$-belief turns out to be $r_{i}$ (in a standard countable-partition model, this is a consequence of a version of the \textquotedblleft sure-thing principle'' (e.g., \cite{Bacharach})). On the other hand, the prior probability in $j$'s $p$-belief in the common $p$-belief conditional on $i$'s $p$-belief in the common $p$-belief is at least $p$. These facts imply $|r_{i} - r_{j}| \leq 1-p$. This also suggests that the failure of agreeing-to-disagree in an infinite non-partitional structure by \cite{SametIJGT} can be seen as the failure of Invariance. 

\subsection{Uniqueness of a Partition Consistent with Quantitative Beliefs}\label{sec:uniquepartition}

In a quantitative belief model $\langle \Omega, \Sigma, \mu, t \rangle$, an information set at a state is often externally introduced as the support of the type at that state in the literature (e.g., \cite{BattigalliBonannoRIE, BNCPA, HalpernKBC, TanWerlangJET, VassilakisZamir} in various settings). In a discrete model, it means that the agent considers $\omega'$ possible (according to her type mapping $t$) at state $\omega$ if $t(\omega, \{ \omega' \})>0$. 

The next corollary shows that if there is a partition $\{ P(\omega) \}_{\omega \in \Omega}$ such that the model $\langle \Omega, \Sigma, \mu, P, t \rangle$ is regular, then $P(\cdot) = [t(\cdot)]$. Roughly, if the analysts would like to introduce an agent's fully introspective knowledge  (i.e., knowledge introduced by a partition) in a quantitative belief model $\langle \Omega, \Sigma, \mu, t \rangle$, then the unique possibility correspondence $P(\cdot)$ which makes the resulting model regular is  $[t(\cdot)]$. The uniqueness part provides a justification for introducing an information set by the support of a type in a discrete model: $P(\omega) = [t(\omega)] = \{ \omega' \in \Omega \mid t(\omega, \{ \omega' \})>0 \}$.

\begin{cor}\label{cor:regular}
\begin{enumerate}
\item Let $\overrightarrow{\Omega}$ be a model. The following are equivalent.
\begin{enumerate}
\item \label{itm:corregulara} $\{ P(\omega) \}_{\omega \in \Omega}$ is a partition; Invariance; Entailment; Self-Evidence of Beliefs; and each $t(\omega, \cdot)$ is a (countably) additive probability measure.
\item \label{itm:corregularb} $P(\cdot) = [t(\cdot)]$ and $t(\cdot, \cdot) = \mu (\cdot \mid P(\cdot))$.
\end{enumerate}
\item Let $\overrightarrow{\Omega}$ be a model satisfying Invariance, Entailment, and Self-Evidence of Beliefs. Suppose further that each $t(\omega, \cdot)$ is a (countably) additive probability measure. Then, $\{ P(\omega) \}_{\omega \in \Omega}$ is a partition (i.e., $K$ satisfies Truth Axiom, Positive Introspection, and Negative Introspection) iff $P(\cdot) = [t(\cdot)]$.
\item Let $\overrightarrow{\Omega}$ be a model such that $P(\cdot) = [t(\cdot)]$ and that each $t(\omega, \cdot)$ is a (countably) additive probability measure. The model satisfies Entailment and Invariance iff $t(\cdot, \cdot) = \mu (\cdot \mid P(\cdot))$.
\end{enumerate}
\end{cor}

The second and third parts of Corollary \ref{cor:regular} follow from the first. The second part establishes the uniqueness of the possibility correspondence $P$ compatible with the consistency conditions. If the analysts introduce knowledge together with quantitative beliefs in a consistent way, the possibility correspondence $P(\cdot) = [t(\cdot)]$ is uniquely determined. The third part states that, under $P(\cdot) = [t(\cdot)]$, the \textquotedblleft Bayes conditional property'' (i.e., $t(\cdot, \cdot) = \mu (\cdot \mid P(\cdot))$) characterizes Entailment and Invariance.

\subsection{Almost-Sure Truth Axiom of Probability-One and Qualitative Beliefs}\label{sec:TA}

Lastly, I compare the qualitative belief and probability-one belief operators in a regular model. It is theoretically important to understand the differences between qualitative and probability-one beliefs. Here, I examine these possible differences in terms of belief operators. First, as is known in the literature (e.g., \cite{MondererSamet, VassilakisZamir}), it is not necessarily the case that $B^{1} = K$ even $\mu$-almost surely. Second, on a related point, while the probability-one belief operator satisfies Positive Introspection and Negative Introspection ($B^{1}(\cdot) \subseteq B^{1}B^{1}(\cdot)$ and $(\neg B^{1}) \subseteq B^{1}(\neg B^{1})(\cdot)$) by Certainty of Beliefs, it is not necessarily the case in a regular model that the qualitative belief operator $K$ satisfies both introspective properties $\mu$-almost surely. Examples \ref{ex:boperators} and \ref{ex:introspection} in Appendix \ref{sec:appendix} demonstrate these two facts. Yet, I show below that the probability-one belief and qualitative belief operators satisfy Truth Axiom $\mu$-almost surely. In a regular interactive epistemic model, this implies that the common $1$-belief and qualitative common belief operators also satisfy Truth Axiom $\mu$-almost surely.

\begin{cor}\label{cor:mainTA}
\begin{enumerate}
\item In any regular model, $B^{1}$ satisfies Truth Axiom $\mu$-almost surely: $\mu ( B^{1}(E) \setminus E) =0$ for all $E \in \Sigma$. Consequently, $K$ also satisfies Truth Axiom $\mu$-almost surely. Indeed, $B^{1}$ and $K$ satisfy Truth Axiom $t(\omega, \cdot)$-almost surely for all $\omega \in \Omega$.
\item In any regular interactive epistemic model, the common qualitative belief operator $C$ and the common $1$-belief operator $C^{1}$ satisfy Truth Axiom $\mu$-almost surely (and consequently, $t(\omega, \cdot)$-almost surely for all $\omega \in \Omega$). 
\end{enumerate}
\end{cor}

Two technical remarks on the first part of Corollary \ref{cor:mainTA} are in order. First, in comparison with Corollary \ref{cor:main}, in a regular discrete model, $B^{1}=K$ satisfies Truth Axiom, Positive Introspection, and Negative Introspection. The first part of Corollary \ref{cor:mainTA} states that, generally, the $1$-belief and qualitative belief operators satisfy Truth Axiom $\mu$-almost surely in any regular model. Second, to obtain almost-sure Truth Axiom of $B^{1}$, it is enough for a given model to satisfy Invariance,  Certainty of Beliefs, and  $\mu([t(\cdot)])>0$. Thus, this part of Corollary \ref{cor:mainTA} also holds in a quantitative belief model $\langle \Omega, \Sigma, \mu, t \rangle$.

A closely-related result is obtained in \cite[Property P.4.]{BDprob1}, where an agent's type mapping is introduced as a posterior conditional on a sub-$\sigma$-algebra that dictates her information. The  probability-one belief operator satisfies Truth Axiom $t(\omega, \cdot)$-almost surely for all $\omega \in \Omega$ ($\mu$-almost surely as well). Also, \cite[Proposition 4.3]{HalpernKBC} establishes that $B^{1}$ satisfies Truth Axiom $\mu$-almost surely when the type mapping does not depend on states (i.e., $t(\omega, \cdot) = t(\omega', \cdot)$ for all $\omega, \omega' \in \Omega$).\footnote{Invariance implies that if the type mapping is independent of states then $\mu(\cdot) = t(\omega, \cdot)$ for all $\omega \in \Omega$.}

\section{Concluding Remarks}\label{sec:conclude}

This paper studied implications of the consistency conditions among prior, posteriors, and information sets on introspective properties of qualitative belief induced from information sets. The consistency conditions are: (i) the prior belief is equal to the expectation of the posterior beliefs (Invariance); (ii) qualitative belief entails probability-one belief (Entailment); and (iii) qualitative belief in one's own quantitative beliefs (Self-Evidence of Beliefs). The main benchmark result (Theorem \ref{thm:main}) states that a model satisfies the consistency conditions iff the information sets form a partition almost surely and the posteriors coincide with the Bayes conditional probabilities given the information sets.

Implications were as follows. First, the posterior at each state is uniquely determined (Corollary \ref{cor:uniquetype}). Second, in a discrete model, the information sets necessarily form a partition (Corollary \ref{cor:main}). I discussed its implications when the information sets do not form a partition, i.e., for qualitative belief violating Truth Axiom or non-introspective knowledge violating Negative Introspection (Corollaries  \ref{cor:unawareness} and \ref{cor:ck}). 

Third, to introduce fully introspective knowledge in a quantitative belief model, the partition generated by the type mapping is a unique partition compatible with the consistency conditions (Corollary \ref{cor:regular}). This result justifies the definition of an information partition by the support of each type in the previous literature. Forth,  while qualitative and probability-one beliefs may differ, both satisfy Truth Axiom almost surely (Corollary \ref{cor:mainTA}). 

Propositions \ref{prop:pcertainty} and \ref{prop:operatorprop} also studied how the additivity of types plays a role in negative introspection of beliefs. As avenues for future research, it is interesting to scrutinize a link between prior and posteriors (or an \textquotedblleft updating rule'') that is consistent with non-partitional information processing. In so doing, it is also interesting to explore the role of additivity. It would also be interesting to study the relation among prior, posteriors, and information sets in the context of a generalized state space model of unawareness.

\appendix

\section{Appendix}\label{sec:appendix}

\begin{exl}\label{ex:boperators}
As in \cite[p.176]{MondererSamet}, suppose that the agent is reasoning about the realization of a random draw from $[0,1]$. Let $\langle \Omega, \Sigma, \mu \rangle = \langle [0,1], \mathcal{B}_{[0,1]}, \mu \rangle$, where $\mu$ is the Lebesgue measure. Let $P(\cdot)=[0,1]$, i.e., the agent considers every number possible at each realization. Her qualitative belief reduces to (degenerate) knowledge in that she only knows that the draw is from $[0,1]$ at each state. Her type at each $\omega$ is $t(\omega, \cdot) = \mu (\cdot)$. By construction, the model is regular. At any realization, the agent does not know that the draw is an irrational number, as she does not observe the realization.  She, however, believes with probability one that the draw is an irrational number. In fact, she $1$-believes any event $E$ at any state $\omega$ as long as $\mu(E)=1$.  For any $E \in \Sigma \setminus \{ \Omega \}$ with $\mu(E) =1$, $B^{1}(E) \setminus K(E) = \Omega$ and $\mu( B^{1}(E) \setminus K(E) ) = 1$. While the agent's qualitative belief is fully introspective knowledge, probability-one belief and knowledge differ. While $K$ satisfies Truth Axiom, $B^{1}$ violates it. Also, $B^{1}$ fails the strong conjunction property that $K$ possesses: for any  $\mathcal{E} \in \mathcal{P}(\Sigma)$ with $\bigcap \mathcal{E} \in \Sigma$, $\bigcap_{E \in \mathcal{E}} K(E) \subseteq K(\bigcap \mathcal{E})$.
\end{exl}

\begin{exl}\label{ex:introspection}
As in Example \ref{ex:boperators}, suppose that the agent is reasoning about the realization of a random draw on the probability space $\langle \Omega, \Sigma, \mu \rangle = \langle [0,1], \mathcal{B}_{[0,1]}, \mu \rangle$. At each realization $\omega \in [\frac{1}{2}, 1] \cup [0,1]_{\mathbb{Q}}$, she considers $P(\omega) = [0,1] \setminus \mathbb{Q}$ possible. At each $\omega \in [0, \frac{1}{2}) \setminus \mathbb{Q}$, she considers $P(\omega) = [0,1]$ possible. Her type at each $\omega$ remains unchanged: $t(\omega, \cdot) = \mu(\cdot)$. The model is regular. Her qualitative belief violates all of Truth Axiom, Positive Introspection, and Negative Introspection. While Truth Axiom holds $\mu$-almost surely (as a consequence of Corollary \ref{cor:mainTA}), the introspection properties do not even in this sense. For example, if $E = [0,1] \setminus \mathbb{Q}$ then $\mu( K(E) \setminus KK(E) ) = \frac{1}{2}$ and $\mu( (\neg K)(E) \setminus K(\neg K)(E) ) =\frac{1}{2}$. 
\end{exl}

\bibliographystyle{eptcs}
\bibliography{BeliefTARKProceeding}

\end{document}